\documentclass{elsart}
\usepackage{graphicx}
\usepackage{amsmath}
\usepackage{amsfonts}
\usepackage{amssymb}
\usepackage{epsfig}
\usepackage{cite}
\usepackage{times}
\usepackage{calc}
\usepackage{version}
\usepackage[french,english]{babel}

\usepackage{epsf}
\usepackage{graphics}
\usepackage{ulem}
\newcommand{\pt}{p_{_\perp}}
\newcommand{\kt}{k_{_\perp}}
\newcommand{\alphas}{\alpha_s}
\newcommand{\sqrtsnn}{\sqrt{s_{_{\mathrm{NN}}}}}
\newcommand{\dd}{{\rm d}}

\def\cO#1{{{\cal{O}}}\left(#1\right)} 
 
\newcommand{\nc}{N_{_c}}
\newcommand{\cf}{C_{_F}}
\newcommand{\tr}{T_{_R}}
\newcommand{\nf}{n_{f}}
\newcommand{\ns}{N_{s}}
\newcommand{\calp}{{\cal P}}
\newcommand{\lqcd}{\Lambda_{_{\rm QCD}}}
\newcommand{\pp}{$p$--$p$\ }
\newcommand{\epem}{e^+e^-}

\begin{document}

\vspace{-2.cm}
\begin{flushright}
\sffamily{LAPTH-1355/09}\\
\sffamily{FTUV-2010}
\end{flushright}

\begin{frontmatter}
\title{Collimation of average multiplicity in QCD jets}

\author[lapth]{Fran\c{c}ois Arleo}\footnote{{\it Email address:} 
\texttt{arleo@lapp.in2p3.fr}}, 
\author[munich,ific]{Redamy P\'erez Ramos}\footnote{{\it Email address:} \texttt{redamy.perez@uv.es}}

\address[lapth]{Laboratoire d'Annecy-le-Vieux de Physique Th\'eorique (LAPTH), \\UMR~5108, Universit\'e de Savoie, CNRS, BP 110, 74941 Annecy-le-Vieux cedex, France}
\address[munich]
{Max Planck Institut f\"ur Theoretische Physik 
(Werner-Heinsenberg-Institut)\\
F\"ohringer Ring 6, 
D-80805, Munich, Germany}
\address[ific]
{Departament de F\'isica Te\`orica and IFIC, Universitat 
de Val\`encia-CSIC,\\
Dr. Moliner 50, E-46100 Burjassot, Spain}

\begin{abstract}
The collimation of average multiplicity inside quark and gluon jets is investigated in perturbative QCD in the modified leading logarithmic approximation (MLLA). The role of higher order corrections accounting for energy conservation and the running of the coupling constant leads to smaller multiplicity collimation as compared to leading logarithmic approximation (LLA) results. The collimation of jets produced in heavy-ion collisions has also been explored by using medium-modified splitting functions enhanced in the infrared sector.  As compared to elementary collisions, the angular distribution of the jet multiplicity is found to broaden  in QCD media at all energy scales.
\end{abstract}
\begin{keyword}
Perturbative QCD, Jets, Multiplicity, Quark-gluon plasma
\end{keyword}
\end{frontmatter}

\setcounter{footnote}{0}
\renewcommand{\thefootnote}{\arabic{footnote}} 	

\section{Introduction}

Paradoxically, the ``jet quenching'' phenomenon observed at RHIC  precedes by many years the first measurements of jets in heavy-ion collisions. The spectacular suppression of large-$\pt$ single-inclusive pion production in central Au--Au collisions has been widely interpreted as coming from the energy loss of hard partons in a dense medium (see e.g.~\cite{dEnterria:2009am}). However, these data --~as well as more differential measurements such as di-hadron and photon-hadron correlations~-- only hardly inform us on quark and gluon multiple scattering processes and medium-induced gluon radiation in quark-gluon plasma (for a review on the subject, we refer the reader to~\cite{Peigne:2008wu}). Measuring hadronic distributions inside reconstructed jets in heavy-ion collisions should therefore be key in order to access the underlying dynamics of jet quenching. Using recent advances in jet algorithm techniques~\cite{Salam:2009jx}, preliminary measurements on inclusive jet spectra have been reported lately in Au--Au collisions at $\sqrtsnn=200$~GeV by the STAR collaboration~\cite{Salur:2009vz}. On the theoretical side, various jet observables have been investigated such as the angular distribution of jet average multiplicities~\cite{Baier:1999ds,Salgado:2003rv}, multiplicity distributions~\cite{Dremin:2006da,Armesto:2008qe} or inclusive momentum spectra~\cite{Borghini:2005em}. The appearance of various parton showers in heavy-ion collisions~\cite{Zapp:2008gi,Renk:2008pp,Armesto:2009fj} should also allow for a more systematic exploration of jet observables and their medium-modifications. 

As for any hard process in heavy-ion collisions, the precise knowledge of the expected baseline in \pp collisions becomes crucial in order to properly quantify the effects of quark-gluon plasma formation on jet physics. Over the past twenty years, significant progresses have been achieved  in order to improve predictions on various jet observables. In particular, the modified leading logarithmic approximation (MLLA) in perturbative QCD has been successfully tested from $\epem$ to hadronic collisions\footnote{For rather exclusive observables such as the transverse momentum of hadrons inside jets, the disagreement between data and MLLA expectations can be cured by the inclusion of higher-order terms, $\cO{\alphas}$, in the calculation~\cite{Arleo:2007wn}.}~\cite{Khoze:1996dn}. 

To this sake, we investigate in this paper the collimation of the average multiplicity inside a quark and a gluon jet as a function of the jet energy scale. The collimation is characterized by the cone aperture of the sub-jet  $\Theta_\delta$ that contains a fraction $\delta$ of the jet average multiplicity. Let $N(E, \Theta_0)$ be the multiplicity in a jet of energy $E$ and opening angle $\Theta_0$, and $\hat N(\Theta; E, \Theta_0)$ that of  the sub-jet of opening angle $\Theta$ inside this jet, $\Theta_\delta$ is determined by solving the multiplicity collimation equation,
\begin{equation*}\label{eq:col}
\hat N(\Theta_\delta; E, \Theta_0)= \delta \times N(E, \Theta_0).
\end{equation*}
 In the leading logarithmic approximation (LLA)~\cite{Dokshitzer:1991wu}, the solid angle $\Theta_{1/2}$ containing half of the average multiplicity of the jet $\Theta_0$ decreases with the jet hardness $Q\simeq E\Theta_0$ approximately like $\Theta_{1/2}(Q)\sim N^{-1/4}(Q)$, such that at high energy scales the bulk of the total multiplicity is concentrated at smaller solid angles around the direction of propagation of the jet. In this paper, we extend this calculation to the MLLA by incorporating all corrections of order $\cO{\sqrt{\alpha_s}}$, which partially guarantee energy conservation and account for the running of the coupling constant $\alpha_s$ in intra-jet cascades. The treatment of such a complicated task in perturbative QCD simplifies tremendously thanks to the angular ordering of the successive emission of soft gluons, which leads to simplified jet evolution equations~\cite{Dokshitzer:1991wu}. At the end of the cascading process, the collinear cut-off parameter $Q_0$ can be taken as low as $\lqcd$ (the so-called limiting spectrum approximation), and the local parton-hadron duality (LPHD) hypothesis~\cite{Azimov:1985by} can then be advocated so as to determine hadronic distributions. 

Within the same formalism, the collimation of jets has also been determined in heavy-ion collisions by inserting medium-modified splitting functions, in which the soft gluon emission is arbitrarily enhanced from a toy QCD-inspired model~\cite{Borghini:2005em}. Although more realistic approaches treating parton energy loss in medium-modified fragmentation have been proposed~(see e.g.~\cite{Arleo:2008dn} for a review), the use of this model was motivated by the fact that analytic calculations can more easily be performed.
 
The outline of the paper is as follows. In Section~\ref{se:framework}, the calculation of the average multiplicity in a sub-jet $\hat N$ is carried out at MLLA, which eventually allows for solving numerically the above collimation equation. In Section~\ref{se:results}, our results are compared to the LLA predictions in the vacuum and medium effects on jet collimation are discussed.

\section{Theoretical framework}\label{se:framework}

\subsection{Energy-multiplicity correlation and sub-jet average multiplicity}

Consider a jet produced in a high energy collision, initiated by a parton of flavour $A_0$ and energy $E$, with an opening angle $\Theta_0$ which separates it from other jets. Inside this jet, a sub-jet initiated by a parton $A$ and energy $u E$ is defined by the opening angle $\Theta<\Theta_0$. The sub-jet multiplicity $\hat N_{A_0}^h$, i.e. the mean number of hadrons produced inside the angular
range $\Theta<\Theta_0$ of the jet $A_0$, 
is given by~\cite{Dokshitzer:1991wu,Ochs:2008vg}
\begin{equation}\label{eq:Nconv}
\hat N_{A_0}^h(\Theta;E,\Theta_0)\approx\sum_{A=q,g}\int_{Q_0/E\Theta}^1 \dd u\,u\,
D_{A_0}^A\left(u,E\Theta_0,uE\Theta\right)
N_A^{h}\left(uE\Theta,Q_0\right),
\end{equation}
which determines the correlation between the jet axis and the average
multiplicity of the sub-jet. 
In Eq.~(\ref{eq:Nconv}), $D_{A_0}^A$ is the probability to find a parton $A$ with energy fraction $u$ and virtuality $u E \Theta\geq Q_0$ inside the jet initiated by the parton $A_0$, and $N_A^h(uE\Theta, Q_0)$ is the bare average multiplicity in a sub-jet of energy $uE$ and opening angle $\Theta$. Since soft particles are less sensible to the energy balance in intra-jet cascades, the correlation between the energy flux and the sub-jet multiplicity disappears. As a consequence, the measured average multiplicity $\hat{N}^h_{A_0}$ factorizes in the form~\cite{Ochs:2008vg}
\begin{equation}\label{eq:Ncc}
\hat N_{A_0}^h(Y_{\Theta_0},Y_{\Theta})
\approx\frac{1}{\nc}\langle C\rangle_{_{A_0}}(Y_{\Theta_0},Y_\Theta)\,
N_g^h(Y_{\Theta}),
\end{equation}
after the Taylor expansion of $D_{A_0}^A$ in Eq.~({\ref{eq:Nconv}}), where we define
\begin{equation*}
Y_{\Theta_0}=\ln\frac{E\Theta_0}{Q_0},\quad Y_\Theta=\ln\frac{E\Theta}{Q_0}.
\end{equation*}
In Eq.~(\ref{eq:Ncc}), $\langle C\rangle _{A_0}$ is the colour current of partons caught by the calorimeter and $N^h_g(Y_{\Theta})$ is the bare average multiplicity  of a gluon jet of energy $E$ and opening angle $\Theta$, given by the solution of the MLLA evolution equations. The colour current $\langle C\rangle _{A_0}$ describes the evolution of the jet between the scales
$Q_{\Theta_0}=E\Theta_0$ and $Q_{\Theta}=E\Theta$ and indicates 
that the registered partons lose
the memory of the initial colour state of the parent parton $A_0$ because of intra-jet evolution~\cite{Dokshitzer:1991wu,Ochs:2008vg}. Introducing the gluon to quark jet total multiplicity, $r=N^h_g/N^h_q$,
the colour current can be written at MLLA accuracy as~\cite{Ochs:2008vg}
\begin{eqnarray}\notag
\langle C\rangle_{_{A_0}}(\xi)&\approx&\nc 
\left[\langle u\rangle_{_{A_0}}^g(\xi)+r^{-1}\ \langle u\rangle_{_{A_0}}^q(\xi)
\right.\\
&+&  \left.\left(\langle u\ln u\rangle_{_{A_0}}^g(\xi)
+r^{-1}\ \langle u\ln u\rangle_{_{A_0}}^q(\xi)\right)
\frac{\dd\ln N_g^h}{\dd Y_{\Theta}}
+{\cal O}(\alpha_s)\right]\label{eq:ccfirst},
\end{eqnarray}
where for the sake of brevity, we introduced the variable\footnote{It should not be confused with the notation $\xi=\ln(1/x)$ used e.g. in~\cite{Ellis:1991qj}.}
\begin{equation}
\xi(Y_{\Theta_0},Y_\Theta)=\frac1{4\nc\beta_0}\ln
\left(\frac{Y_{\Theta_0}}{Y_\Theta}\right),
\end{equation}
with $\nc=3$ is the number of colours, $\beta_0=\frac1{4\nc}\left(\frac{11}3\nc-\frac43\tr\right)$
is the first coefficient of the QCD $\beta$-function and $\tr=\nf/2$
where $\nf=3$ is the number of active flavours. 
The functions appearing in Eq.~(\ref{eq:ccfirst}) are written in the form
\begin{subequations}\label{eq:ulogu}
\begin{eqnarray}
\langle u\rangle_{A_0}^{A}(\xi)&\equiv&\int_0^1du\ u\
D_{A_0}^{A}(u,\xi)=
{\cal D}_{A_0}^A(j=2,\xi),\\
\langle u\ln u\rangle_{A_0}^{A}(\xi)&\equiv&\int_0^1du\ u\ \ln u\
D_{A_0}^{A}(u,\xi)=\frac{d}{dj}
{\cal D}_{A_0}^A(j,\xi)\Big|_{j=2},
\end{eqnarray}
\end{subequations}
which can be determined from the DGLAP evolution equations~\cite{Ochs:2008vg}.

\subsubsection{Medium-modified DGLAP and MLLA evolution}

In order to compute Eq.~(\ref{eq:ccfirst}) from Eq.~(\ref{eq:ulogu}), 
the moments ${\cal D}(j, \xi)$ need first to be 
determined by solving the DGLAP equation in Mellin space,
\begin{equation}\label{eq:dglap}
\frac{d}{d\xi}
\begin{pmatrix}
{\cal D}_{q_{\rm ns}}(j,\xi) \\
{\cal D}_{q_{\rm s}}(j,\xi) \\
{\cal D}_{g}(j,\xi)
\end{pmatrix} 
=\begin{pmatrix}
\calp_{qq}(j)&0&0 \\
0&\calp_{qq}(j)&\calp_{q g}(j)\\
0&\calp_{gq}(j)&\calp_{gg}(j) 
\end{pmatrix}
\begin{pmatrix}
{\cal D}_{q_{\rm ns}}(j,\xi) \\
{\cal D}_{q_{\rm s}}(j,\xi) \\
{\cal D}_{g}(j,\xi)
\end{pmatrix},
\end{equation}
where ${\cal D}_{q_{\rm ns}}$ and ${\cal D}_{q_{\rm s}}$ stand respectively for the flavour non-singlet (or valence) 
and flavour-singlet quark distributions, and $\calp_{ik}(j)$ 
is the Mellin transform of the leading-order splitting functions
$P_{ik}(z)$. In order to account for the medium-induced gluon 
radiation in heavy-ion collisions, various attempts  to determine 
medium-modified splitting functions have been recently performed~\cite{Armesto:2007dt,Domdey:2008gp,Nayak:2009iw}. In this 
paper, we shall adopt the most simple approach proposed by Borghini 
and Wiedemann~\cite{Borghini:2005em} which allows for analytic 
solutions. In this model the infrared sensitive parts of the splitting 
functions are enhanced by a factor\footnote{In~\cite{Borghini:2005em} 
the parameter $f_{_{\rm med}}$ is used, corresponding to $\ns-1$ here.} 
$\ns$,
\begin{subequations}
\begin{equation}\label{eq:splitG}
P_{gg}(z)=4\nc\left[\frac{\ns}{z}+\left[\frac{\ns}{1-z}\right]_+
+z(1-z)-2\right],\quad
P_{gq}(z)=2\ \tr[z^2+(1-z)^2],
\end{equation}
\begin{equation}\label{eq:splitQ}
P_{q g}(z)=2\ \cf\left(\frac{2\ns}{z}+z-2\right),\quad
P_{qq}(z)=2\ \cf\left(\left[\frac{2\ns}{1-z}\right]_+-1
-z\right),
\end{equation}
\end{subequations}
with the $[\dots]_+$ prescription defined as $\int_0^1dx[F(x)]_+g(x)\equiv\int_0^1dxF(x)[g(x)-g(1)]$. Performing the Mellin transform of Eq.~(\ref{eq:splitG},\ref{eq:splitQ}) gives~\cite{Albino:2009hu}
\begin{subequations}\label{eq:splitmellin}
\begin{eqnarray}
\calp_{gg}(j)&=&-4 \nc\left[\ns\ \psi(j+1)+\ns\gamma_E-\frac{\ns-1}{j}-\frac{\ns-1}{j-1}\right]\nonumber\\ 
&&+\frac{11 \nc}3-\frac{2\nf}3+\frac{8\nc(j^2+j+1)}{j(j^2-1)(j+2)}\label{eq:nuGbis},\\
\calp_{gq}(j)&=&\tr\ \frac{j^2+j+2}{j(j+1)(j+2)},\label{eq:splitjG} \\
\calp_{q g}(j)&=&2\ \cf\
\frac{(2 \ns-1)(j^2+j)+2}{j(j^2-1)}\label{eq:splitjQ},\\
\calp_{qq}(j)&=&-\cf\left[4\ns\ \psi(j+1)+4\ns\gamma_E-4\frac{\ns-1}j-3-\frac2{j(j+1)}\right]\label{eq:nuFbis},
\end{eqnarray}
\end{subequations}
where $\psi(j)$ is the digamma function. It can be easily checked
that Eq.~(\ref{eq:splitmellin}) reduces to the ordinary splitting functions given in~\cite{Dokshitzer:1991wu,Dokshitzer:1978hw} after setting $\ns=1$. 
Using Eq.~(\ref{eq:splitmellin}), the DGLAP equation 
Eq.~(\ref{eq:dglap}) can be solved, such that the expressions for
${\cal D}_q^q(j,\xi)={\cal D}_{q_{\rm ns}}(j,\xi)+{\cal D}_{q_{\rm s}}(j,\xi)$, 
${\cal D}^{g}_q(j,\xi)$, ${\cal D}^{q}_g(j,\xi)$
and ${\cal D}^{g}_g(j,\xi)$ can be taken
from~\cite{Dokshitzer:1991wu} by introducing the $N_s$ dependence from
Eq.~(\ref{eq:splitmellin}).
Finally, these distributions
and their derivatives
can be evaluated at $j=2$ 
in order to determine the moments $\langle u\rangle$ and 
$\langle u \ln u\rangle$, cf. Eq.~(\ref{eq:ulogu}).

The other ingredient that enters the evaluation of the colour currents Eq.~(\ref{eq:ccfirst}) are the MLLA gluon and quark jet average multiplicities~\cite{Dokshitzer:1991wu,Dokshitzer:1978hw, Malaza:1985jd}. 
In the model of~\cite{Borghini:2005em}, they read~\cite{Ramos:2008qb}
\begin{subequations}
\begin{equation}\label{eq:mllaform}
N_g^h(Y_\Theta)={\cal K}\ {Y_\Theta}^{-\sigma_1/\beta_0}
\exp{\sqrt\frac{4\ns\ Y_\Theta}{\beta_0}},
\end{equation}
\begin{equation}\label{eq:mllaformbis}
\frac{d\ln N_g^h}{dY_\Theta}=\sqrt{\ns}\ {\gamma_0}, \quad r=\frac{N_c}{C_F}\left(1-r_1\frac{\gamma_0}{\sqrt{N_s}}\right).
\end{equation}
\end{subequations}
with $\sigma_1\simeq0.28$, and $r_1\simeq0.185$ for $n_f=3$. The anomalous
dimension, $\gamma_0$, determines the rate of the multiplicity increase in a jet; it
can be written
\begin{equation}
\gamma_0^2(E\Theta)=2\nc\ \frac{\alpha_s(E\Theta)}{\pi}=\frac1{\beta_0Y_\Theta}.
\end{equation}
The formul{\ae} in Eqs.~(\ref{eq:mllaform},\ref{eq:mllaformbis}) 
follow from the MLLA evolution
equations as a consequence of angular ordering in the partonic shower. The constant
${\cal K}$ normalizes the number of partons to the number of charged
hadrons according to the LPHD hypothesis~\cite{Azimov:1985by}; as can be checked below, 
this constant does not play any role in this context. Finally,
working out the structure of Eq.~(\ref{eq:ccfirst}), one has
\begin{eqnarray}\label{eq:C2}
\langle C\rangle_{_{A_0}}(\xi)&=&\nc\
\langle u\rangle_{_{A_0}}^g(\xi)+
\frac{1}{r}\ \nc\
\langle u\rangle_{_{A_0}}^q(\xi)\\
&+&\gamma_0 \times \sqrt{\ns}\ 
\left[\nc\langle u\ln u\rangle_{_{A_0}}^g(\xi)+\cf\langle u\ln u\rangle_{_{A_0}}^q(\xi)
\right],\notag
\end{eqnarray}
which will be used in the present form in the following.
As expected, the MLLA expressions of
the colour current in the vacuum~\cite{Ochs:2008vg} are recovered when setting $\ns=1$ in Eq.~(\ref{eq:C2}).

\subsection{Collimation of average multiplicity inside a jet}

As stressed in the Introduction, the collimation is characterized by 
 the angular size $\Theta_\delta$ of the cone containing 
the fraction $\delta<1$ of the total multiplicity in a jet, 
\begin{equation}\label{eq:col1}
\hat N_{A_0}^h(Y_{\Theta_0},Y_{\Theta_\delta})= 
\delta \times N_{A_0}^h(Y_{\Theta_0}).
\end{equation}
Using Eq.~(\ref{eq:Ncc}), the collimation equation 
(\ref{eq:col1}) becomes
\begin{equation}\label{eq:collimation}
\frac{1}{\nc}\ \langle C\rangle_{_{A_0}}(Y_{\Theta_0},Y_{\Theta_\delta})\,
N_g^h(Y_{\Theta_\delta})=\delta \times N_{A_0}^h(Y_{\Theta_0}),
\end{equation}
which can be solved numerically by using the 
analytic expressions for the colour currents and 
the total jet average multiplicity.

The collimation of average multiplicity in a 
jet has been determined in~\cite{Dokshitzer:1991wu}
 at LLA\footnote{Note that the collimation of the {\it energy} 
has also been studied at LLA 
in~\cite{Dokshitzer:1991wu}.}. At large opening angles 
$\Theta_0\sim1$, the LLA collimation was found to scale like
\begin{equation}\label{eq:llacollim}
\Theta_\delta\sim \left[{N_g^h(E/\lqcd)}\right]^{-\frac12\beta_0\ln\frac1{\delta}},
\end{equation}
indicating that the bulk of the jet multiplicity concentrates closer to the jet axis as the energy scale increases. For the sake of more accurate predictions, this result is extended in the present paper to MLLA accuracy, i.e. 
keeping track of all terms of order $\cO{\sqrt{\alpha_s}}$ in~(\ref{eq:collimation}). Moreover, by inserting an explicit $N_s$-dependence in the splitting functions, we provide results accounting for enhanced soft gluon radiation 
in dense QCD media ($N_s>1$).

\section{Phenomenology}\label{se:results}

\subsection{Collimation from LLA to MLLA}

\begin{figure}[ht]
\begin{center}
\epsfig{file=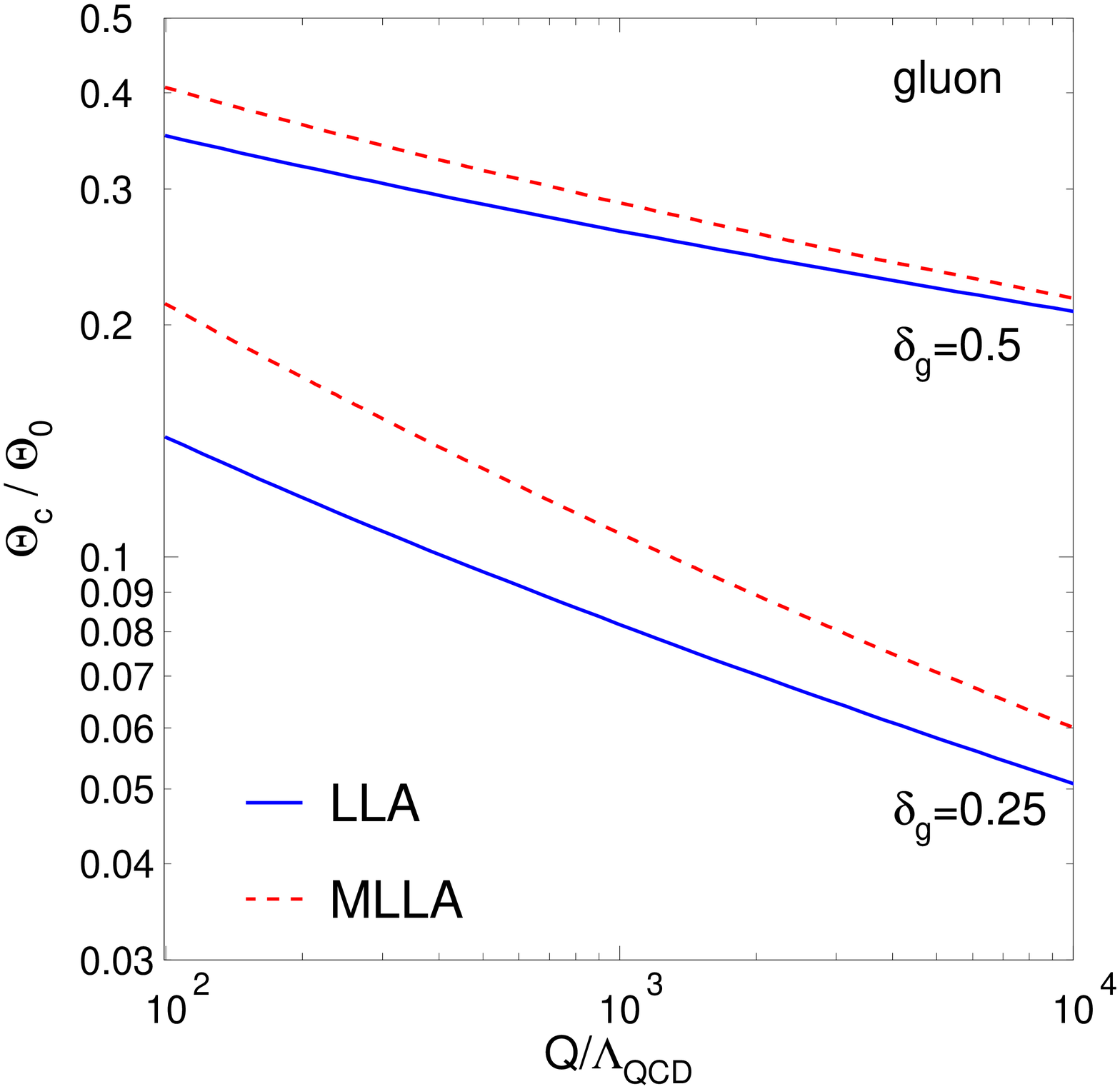,width=6.8truecm}
\epsfig{file=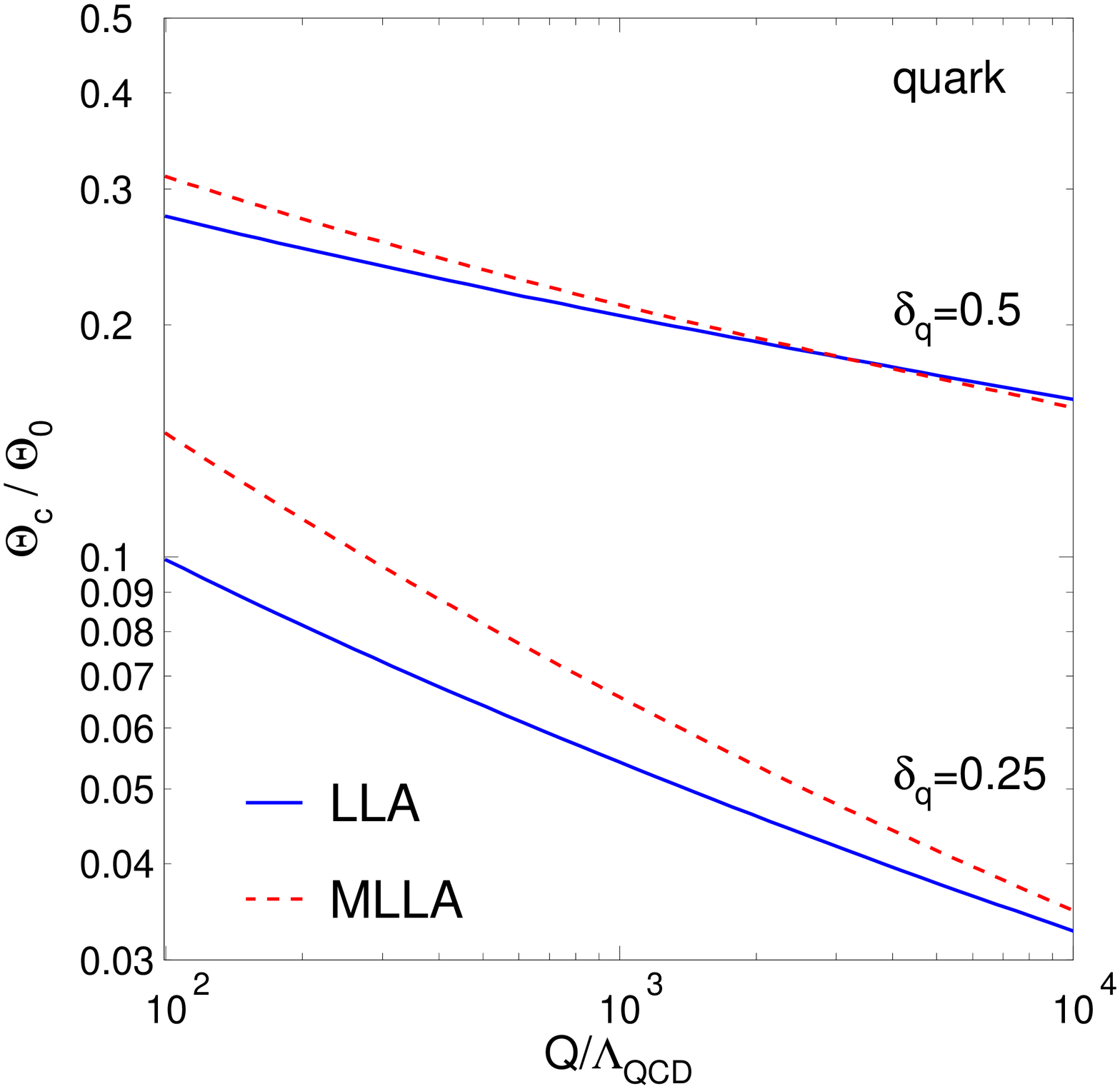,width=6.8truecm}
\caption{Collimation of average multiplicity inside a gluon jet (left)
and a quark jet (right) in the vacuum at LLA (solid) and MLLA (dashed) accuracy.
\label{fig:collim}
}
\end{center}
\end{figure}

In this Section, the multiplicity collimation of jets produced in the vacuum 
is computed at MLLA and compared to the LLA prediction, recovered by setting $\sigma_1$,
$r_1$ and $\gamma_0$ to zero in Eq.~(\ref{eq:collimation}). In Fig.~\ref{fig:collim}, the normalized angular aperture, $\Theta_\delta/\Theta_0$, is plotted as a function of the jet energy scale $Q=E\Theta_0$, in units of $\lqcd$, for gluon (left) and quark (right) jets at LLA (solid lines) and MLLA (dashed) accuracy. For the illustration, $\Theta_\delta$ has been determined using $\delta=0.25$ and $0.5$.

Although the trends are similar, $\Theta_\delta/\Theta_0$ turns out to be larger at MLLA than at LLA, indicating that corrections following from energy conservation and the running of the coupling constant increase the size of the angular distribution, i.e. making jets less collimated at MLLA. As expected, MLLA corrections are larger at low jet energy scales and closer to LLA at asymptotic energies as the anomalous dimension vanishes. Also, MLLA corrections prove slightly larger for a gluon than for quark jet, yet the difference is rather small. More remarkably, the difference between LLA and MLLA predictions is somewhat larger for smaller values of the parameter $\delta$. At low scales, $Q/\lqcd=10^2$, $\Theta_c$ is increased by $50\%$ from LLA to MLLA for $\delta=0.25$, but only by $15\%$ when $\delta$ is set to $0.5$. 

These predictions could be tested at the Tevatron and the LHC, by measuring jets of transverse momenta say $\pt\sim100$~GeV and angular aperture $\Theta_0\sim0.5$, corresponding to energy scales $Q/\lqcd\sim200$. Lower scales, for which MLLA predictions are largest, could even be reached at RHIC energy.

\subsection{Medium effects}

Let us now discuss how the multiplicity collimation of jets produced in heavy-ion collisions could be distorted by quark-gluon plasma formation. Preliminary 
predictions have been made by enhancing the infrared gluon emission by $60\%$ and $80\%$ (respectively, $\ns=1.6$ and $\ns=1.8$) and compared to the MLLA vacuum expectations ($\ns=1$). The values of $\ns$ are somehow arbitrary; these numbers were chosen in~\cite{Borghini:2005em} in order to reproduce the magnitude of single-inclusive pion suppression data in Au--Au collisions at RHIC.

\begin{figure}[ht]
\begin{center}
\epsfig{file=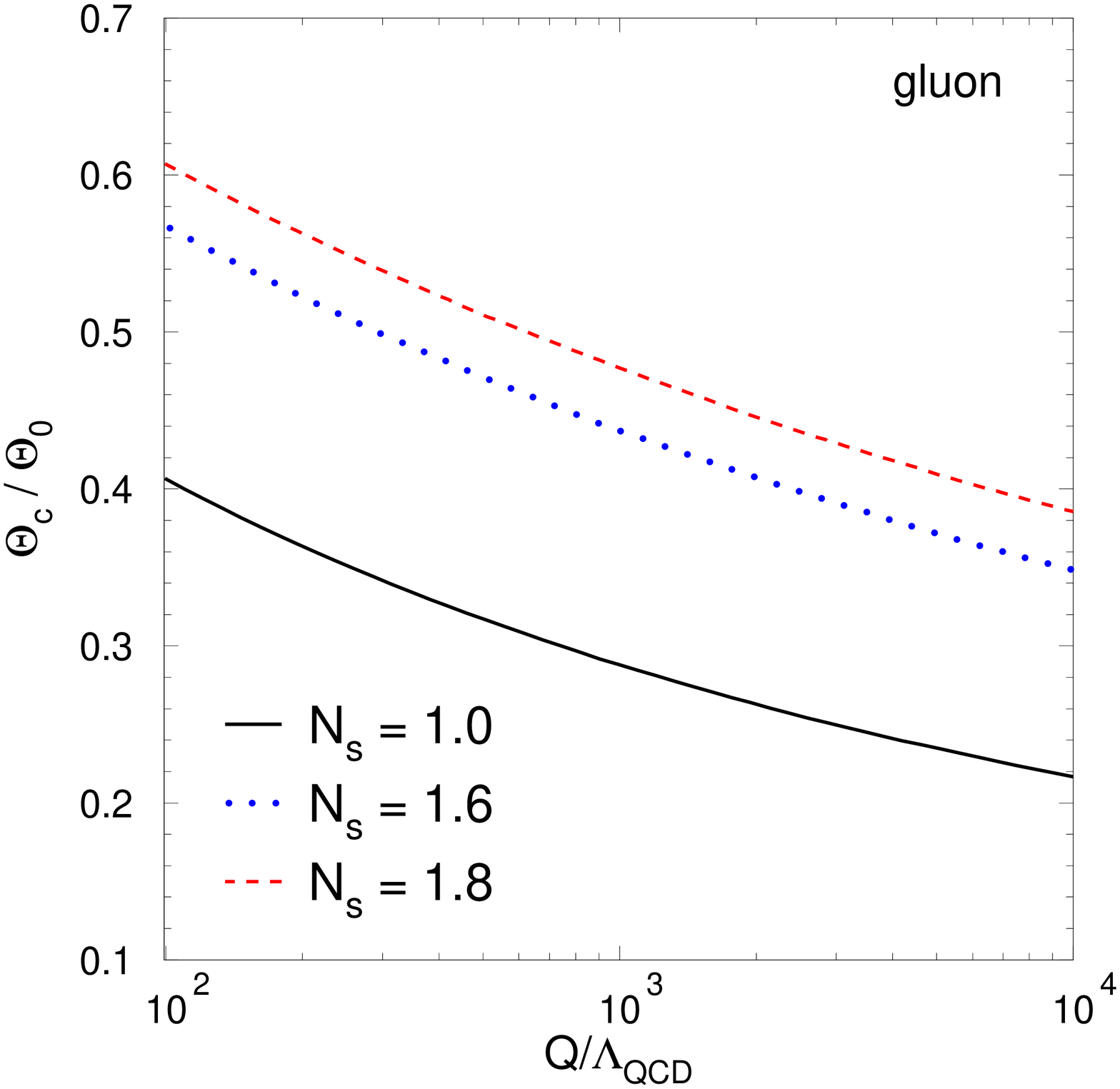,width=6.8truecm}
\epsfig{file=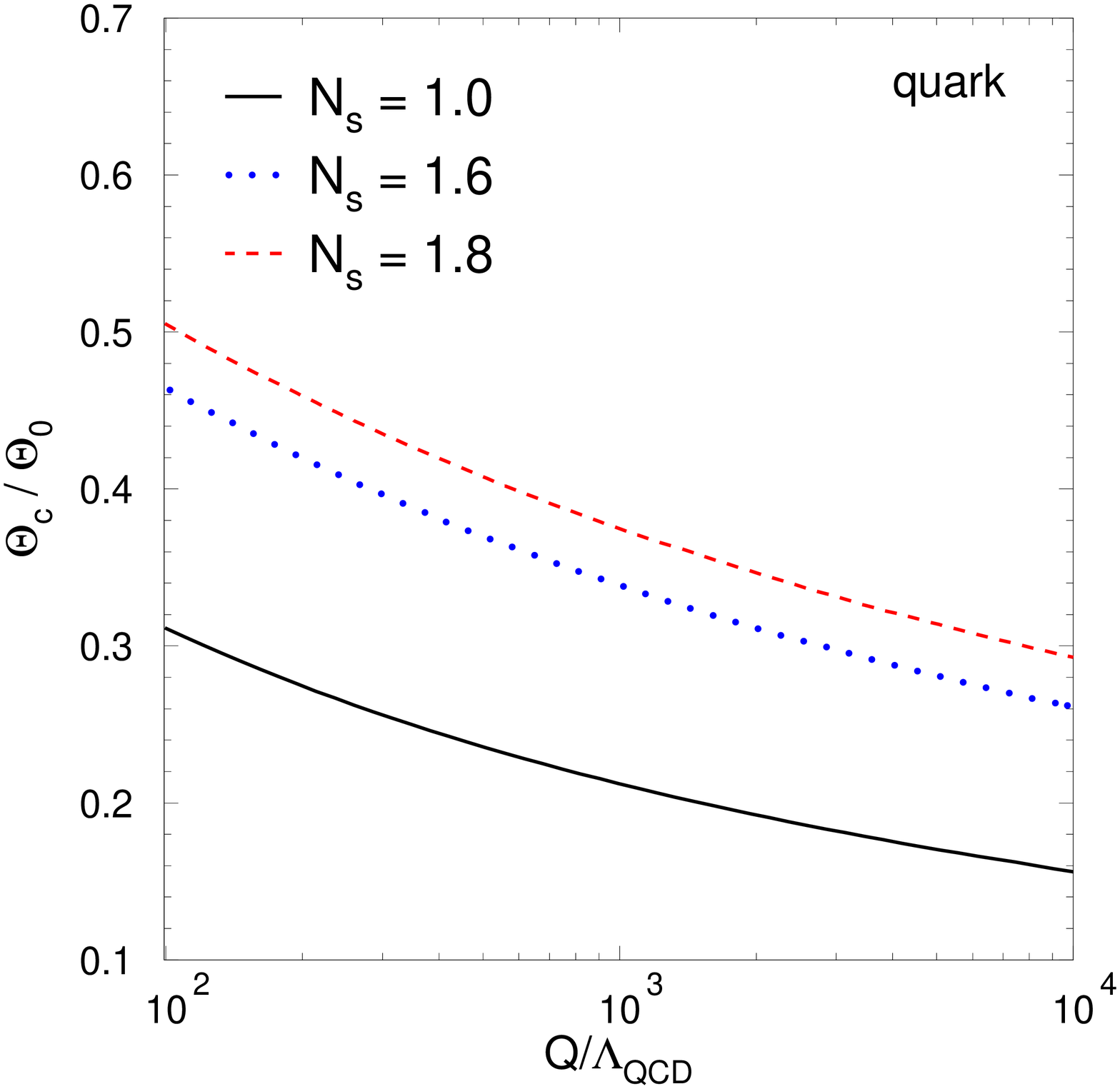,width=6.8truecm}
\caption{Collimation of average multiplicity for $\delta=0.5$ 
inside a gluon jet (left)
and a quark jet (right) produced in the vacuum ($\ns=1$) and in the medium ($\ns=1.6$ and $\ns=1.8$).
\label{fig:collim_med}
}
\end{center}
\end{figure}

The predictions are displayed for gluon (jets) and quark (right) jets in Fig.~\ref{fig:collim_med} by setting $\delta=0.5$. As expected from the role of soft gluon radiation in jet cascades, the angular size $\Theta_\delta$ broadens as compared to the vacuum case.  The typical values for the angular aperture $\Theta_\delta/\Theta_0$ is increased by roughly  factor $0.15$--$0.2$ as compared to jets produced in the vacuum, depending on the values of $Q/\lqcd$ and $\ns$. This observable could be measured in heavy-ion collisions at RHIC and at the LHC provided the background from the underlying event can be safely removed from the jet signal, using the techniques discussed in~\cite{Salam:2009jx}.
On top of an increased of the total multiplicity, see Eq.~(\ref{eq:mllaform}), jets produced in heavy-ion collisions appear to be less collimated than in the vacuum, according to the model \cite{Borghini:2005em} used here. This observation seems qualitatively consistent with the $\kt$-broadening of the jet multiplicity distributions studied in~\cite{Salgado:2003rv}, yet this jet shape is directed related to the collimation of {\it energy} rather than to the collimation of multiplicity considered here. 

The calculations of the medium-modified collimation of multiplicity discussed in this Section are not meant to be quantitative, given the rather primitive model used to describe the process of parton multiple scattering and induced-gluon radiation. Nevertheless, the qualitative prediction of a reduced collimation of jets produced in heavy-ion collisions is expected to hold in more realistic calculations. The procedure presented in this paper in order to compute the collimation of multiplicity at MLLA from given splitting functions could for instance be used in the future to provide more accurate predictions in heavy-ion collisions. 

\section{Summary}

In this paper we have considered the collimation of multiplicity 
distributions inside jets produced at high energy colliders. 
The calculation performed at LLA in~\cite{Dokshitzer:1991wu} 
has been extended to MLLA accuracy by taking full care of all terms of order $\cO{\sqrt{\alpha_s}}$ in the jet evolution. The sub-jet containing a fraction $\delta$ of the total jet multiplicity has been found to widen at MLLA as compared to previous results at LLA. The difference between LLA and MLLA expectations are largest at small jet energy scales and could be tested at present (RHIC, Tevatron) and future (LHC) hadronic colliders.

The same analysis was performed by using the medium-modified splitting funtions introduced in~\cite{Borghini:2005em}, which enhance the soft gluon radiation of a highly virtual parton traveling through quark-gluon plasma. The calculation has also been performed at MLLA and compared to the vacuum expectations. Results indicate that jets produced in heavy-ion collisions are expected to be less collimated than in the vacuum, even though quantitative predictions using more realistic medium-modified splitting functions are yet to be performed.

\section*{Acknowledgments}

We would like to thank M.-A. Sanchis Lozano for useful comments on the manuscript. F.~A. thanks the hospitality of CERN PH-TH department where part of this work has been performed. R.~P.~R. is grateful for the hospitality and nice stay at Max Planck Institut f\"ur Theoretische Physik in Munich and thanks W. Ochs for enlightening discussions. He also thanks the warm welcome at Universitat de Val\`encia-CSIC and IFIC, where this work was completed, and acknowledges support from the Generalitat valenciana under grant PROMETEO/2008/004.

\providecommand{\href}[2]{#2}\begingroup\raggedright\endgroup
\end{document}